\documentstyle[12pt]{article}
\begin{document}
\begin{titlepage}
\begin{center}

{\Large Measuring non-extensitivity parameters in a turbulent
Couette-Taylor flow}

\vspace{2.cm}
{\bf Christian Beck}

\vspace{0.5cm}

School of Mathematical Sciences, Queen Mary and Westfield College,
University of London, Mile End Road, London E1 4NS, UK

\vspace{1cm}

{\bf Gregory S. Lewis}

\vspace{0.5cm}

Calimetrics, Inc., 815 Atlantic Ave., Suite 105, Alameda, California 94501, USA

\vspace{1cm}

{\bf Harry L. Swinney}

\vspace{0.5cm}

Center for Nonlinear Dynamics and Department of Physics, The
University of Texas at Austin, Austin, Texas 78712, USA
\end{center}

\vspace{2cm}

\abstract{We investigate probability density functions of velocity
differences at different distances $r$ measured in a Couette-Taylor
flow for a range of Reynolds numbers $Re$. There is good agreement
with the predictions of a theoretical model based on non-extensive
statistical mechanics (where the entropies are non-additive for
independent subsystems). We extract the scale-dependent
non-extensitivity parameter $q(r,Re)$ from the laboratory
data.}

\vspace{1.3cm}

\end{titlepage}

\section{Introduction}

Recently a generalization of the ordinary formalism of statistical
mechanics, so-called non-extensive statistical mechanics, has
gained a lot of interest \cite{tsa1,tsa2,tsa3}. While in
ordinary statistical mechanics the Boltzmann-Gibbs entropy is
extremized (subject to constraints), in the generalized formalism
the more general Tsallis entropies $S_q$ are extremized. Given
probabilities $p_i$ of the microstates of the physical system
under consideration, the Tsallis entropies are defined as
\begin{equation}
S_q= \frac{1}{q-1} \left( 1- \sum_i p_i^q \right),
\end{equation}
where the parameter $q$ is the non-extensitivity parameter.
The Tsallis entropies are convex, take on their extremum for the
uniform distribution, and preserve the Legendre transform
structure of thermodynamics. However, in contrast to the
Boltzmann-Gibbs entropy, they are non-extensive (non-additive for
independent subsystems) provided $q\not=1$.

Ordinary statistical mechanics is contained as a special case in the
generalized formalism, since in the limit $q\to 1$ the Tsallis
entropies reduce to the Boltzmann-Gibbs entropy, $S_1= -\sum_i p_i\ln
p_i$. Recent work indicates that the non-extensive formalism with
$q\not=1$ describes many systems, including those exhibiting Levy-type
anomalous diffusion \cite{levy}, particles produced in collider
experiments near the Hagedorn phase transition
\cite{curado,parti,qua}, and various turbulent systems
\cite{turb1,turb2,turb3,hydro}. A list of references on non-extensive
statistical mechanics and its applications is given in
\cite{web}. Generally, it is not known how the parameter $q$ depends
on the internal properties of the system under consideration.

In this paper we will apply the non-extensive formalism to turbulence.
In \cite{hydro}, fully developed turbulent states were distinguished
from spatio-temporal chaotic states that extremize the Tsallis
entropies, and formulas were obtained for probability densities of
longitudinal velocity differences measured at a distance $r$.  Here we
test the theoretical predictions by comparing them with experimental
data obtained for turbulent Couette-Taylor flow for different Reynolds
numbers $Re$ and spatial scales $r$.  For details on the experiment,
see \cite{LS}.

We will provide evidence that a slightly generalized version of the
theory described in \cite{hydro} yields a good fit to the
experimentally measured probability densities. These fits were
achieved by varying only one quantity, the non-extensitivity parameter
$q$. We will also present the first systematic experimental
results showing how $q$ depends on $Re$ and $r$.

The probability densities obtained from non-extensive statistical
mechanics asymptotically decay with a power law with a rather large
exponent $w$. In our model $w$ is related to $q$ by
$w=(4-2q)/(1-q)$. We will provide experimental evidence that there is
a simple scaling law for the function $w(r, Re )$.

This paper is organized as follows. Section 2 describes the
experiment.  Section 3 summarizes the theory presented in
\cite{hydro}, and the formalism is slightly generalized.  Section 4 
compares the experimentally measured probability densities
with the theoretical predictions. Section 5 examines the dependence of
the non-extensitivity parameter $q$ and the exponent $w$ on 
Reynolds number and the spatial scale.

\section{Experiment}

The experiments were conducted on a concentric cylinder system with
the inner cylinder rotating and the outer cylinder at
rest \cite{LS}. This Couette-Taylor system had a radius ratio of 0.724.
Measurements were made for Reynolds numbers $Re$ up to 540000, where
$Re=\Omega a(b-a)/\nu$ ($\Omega$ is the inner cylinder rotation rate,
$a$ and $b$ are the inner and outer cylinder radii, and $\nu$ is the
kinematic viscosity). With $Re=0$ initially, the flow exhibits a sequence
of bifurcations with increasing $Re$. The last transition that has
been observed occurs at $Re=13000$ \cite{LS}. Here we consider data for
$Re \geq 69000$, where the flow is strongly turbulent.

Velocity measurements were made with a hot film probe located midway
between the two cylinders. The rms velocity fluctuations were
typically only 6$\%$ of the mean velocity; hence the data satisfy
Taylor's frozen turbulence hypothesis, that is, temporal fluctuations
recorded by the fixed velocity probe should accurately reflect the
streamwise spatial fluctuations \cite{LS}.

\section{Theoretical model for
probability densities of longitudinal velocity differences}

In \cite{hydro} a perturbative approach to probability densities in
fully developed turbulent flows was suggested, based on a small
parameter $\sqrt{\gamma\tau}$, where $\tau$ is a typical time scale of
the chaotic force driving the local velocity differences and
$\gamma^{-1}$ is the relaxation time to the stationary state. From
the experimentally observed skewness of turbulent distributions one
can estimate the order of magnitude of the parameter
$\sqrt{\gamma\tau}$ to be about $0.1$; hence a perturbative approach
makes sense. Assuming that longitudinal velocity differences, $u(r)
\equiv v(x+r)-v(x)$, in fully developed turbulence extremize the
Tsallis entropies, and that large classes of chaotic relaxation
processes approach the Gaussian limit in a universal way (see
\cite{hil}), one can obtain the following formulas \cite{hydro},

\begin{eqnarray}
p(u)&=&\frac{1}{Z_q}\left( 1+\beta (q-1) \epsilon (u)
\right)^{-\frac{1}{q-1}}\label{pu}
\\
\epsilon (u) &=&\frac{1}{2} u^2 -c\sqrt{\gamma\tau} \left(
u-\frac{1}{3}u^3\right) +O(\gamma\tau ) \label{almut}  \\ \beta
&=&\frac{2}{5-3q},
\end{eqnarray}
where $p(u)$ is the stationary probability density of velocity
differences, and $\epsilon (u)$ is a (formal) effective energy
associated with the velocity difference $u$. $\beta$ is a variance
parameter that describes a (formal) inverse temperature in the 
non-extensive statistical mechanics. $Z_q$ is a normalization
constant. $c$ is a non-universal constant, i.e., a constant that may be
different for different experiments and that can also depend on
$q$. However, the functional form $u-\frac{1}{3}u^3$ of the term of
order $\sqrt{\gamma\tau}$ is expected to be universal (see
\cite{hydro, hil}). The parameter $q$ depends on the
distance $r$ in an (a priori) unknown way.

For $c=0$ and $\beta=2/(5-3q)$, the density in eq. (2) has average
value 0 and variance 1. However, if $c\not=0$ and $\beta$ is still
$2/(5-3q)$, then the average of $\langle u \rangle$ is of order
$\gamma\tau$ and the variance $\sigma =(\langle u^2\rangle -\langle
u\rangle^2)^{1/2}$ is slightly different from 1. An average precisely
zero and a variance of unity are achieved with the rescaled
(renormalized) distribution given by

\begin{equation}
 \tilde{p}(u)=\sigma p(\sigma (u-\langle u\rangle).
\end{equation}

The term $\epsilon (u)$ in eq.~(\ref{almut}) stands for an
effective energy in the formalism of non-extensive statistical
mechanics. Indeed, for $q\to 1$, the Boltzmann factor $p(u)\sim
e^{-\beta \epsilon (u)}$ is recovered. Let us here slightly
generalize the approach of \cite{hydro} by considering more
general effective energy levels given by
\begin{equation}
\epsilon(u)=\frac{1}{2} |u|^{2\alpha}-c\sqrt{\gamma\tau} sign (u)
\left( |u|^\alpha-\frac{1}{3} |u|^{3\alpha}\right). \label{ea}
\end{equation}
For $\alpha =1$, eq.~(\ref{almut}) is recovered, but
we will allow for more general exponents $\alpha$ as well. The
physical idea behind this is similar to that of the $\beta$-model
of turbulence \cite{nel}, where only a certain fraction $\beta$ of
the physical volume is considered to contain active eddies.
In the $\beta$-model one essentially replaces the
structure functions $\langle |u|^m \rangle$ by structure functions
$\langle |u|^{\alpha m}\rangle$ with $\alpha \not=1$ related
to the intermittency parameter.
Similarly, in our dynamical model we replace $|u|$ by
$|u|^\alpha$. An exponent $\alpha$ slightly smaller than 1 may be
interpreted as describing a fractal phase space and eddies that
are not space-filling. The formalism of non-extensive statistical
mechanics is designed to include such systems.

In the following we will see that while the choice $\alpha =1$
yields reasonably good fits of the experimental data, an exponent
$\alpha$ slightly smaller than 1 yields the best fits to the
data.

\section{Comparison with the experimental measurements}

We compare the probability density functions determined 
from the Couette-Taylor velocity measurements
with the theoretical densities $\tilde{p}(u)$ obtained from
non-extensive statistical mechanics. The result is shown in
Fig.~1 for Reynolds number $Re =540000$, equivalent to a
Taylor scale Reynolds number $R_\lambda =262$ \cite{LS}. The best
fits were obtained by choosing $\alpha$ according to the empirical
formula $ \alpha =1-(q-1)=2-q$. For the strength of the skewness
term we have chosen, $c\sqrt{\gamma\tau} =0.124(q-1)$. Then only one
independent parameter $q$ is left, which is fitted for each
experimentally measured distribution in such a way that the
relative mean square deviation integrated over $u$ takes a
minimum. Although we only vary a single
parameter $q$, the agreement with the experimentally measured
densities is excellent. If $\alpha$ is chosen as 1 (as
originally suggested in \cite{hydro}), the agreement is still
reasonable but not as good as for $\alpha = 2-q$.

Other theoretical approaches to turbulent densities
\cite{cast}--\cite{casca}, based, for example, on stretched
exponentials or other functional forms, usually fit only certain parts
of the distribution (e.g., the tails), leaving other parts (e.g. the
vicinity of the maximum) unaccounted for. Our formula yields good fits
of the experimental data for the entire range of $u$-values. To
demonstrate the good fit for all $r$ we present both a logarithmic
plot, which emphasizes the tails (Fig.~1(a)), and a linear plot, which
emphasizes the maximum and its vicinity (Fig.~1(b)).  To the best of
our knowledge, there is no other theoretical model that yields fits of
similar quality. Figure 2 shows that the residual difference between
theory and experiments for the same spatial scale, $r/\eta \approx 11$
(where $\eta$ is Kolmogorov scale \cite{LS}), decreases with
increasing $Re$.

The function $q(r,Re)$, deduced from the experiment, will be examined
in the next section.  All relevant information on the densities
appears to be encoded in this function, which is similar to an
equation of state in ordinary thermodynamics.


\section{Measuring $q(r,Re)$}

We have determined the function $q(r,Re )$ for different spatial
scales $r$ and Reynolds numbers $Re$ by minimizing the integrated
relative quadratic deviation between measured distributions and
theoretical distributions. For each Reynolds number, 38 different
spatial scales were evaluated and the results are shown in Fig.~3. The
non-extensitivity parameter $q$ varies with both the spatial scale
$r/\eta$ and with the Reynolds number. There is a tendency to smaller
$q$ for smaller $Re$. For small $r$, $q$ approaches the value
$q_0\approx 1.185$.  For large $r$, $q$ does not approach 1 but
saturates at a slightly larger value, $q_\infty \approx 1.03$, the
precise value being Reynolds number dependent. Hence small deviations
from the Gaussian distribution remain at the largest scales. This may
be a finite-system size effect.

The predicted probability densities for large $|u|$ exhibit power law
decay, as can be deduced from eq.~(\ref{pu}) and (\ref{ea}), where we
neglect the term $O(\sqrt{\gamma\tau })$ (our formula for this term
represents a perturbative result valid only for
$|u|<1/\sqrt{\gamma\tau}\approx 10$):

\begin{equation}
p(u)\sim |u|^{-w}
\end{equation}
with exponent $w$ given by
\begin{equation}
w =\frac{2\alpha}{q-1}=\frac{4-2q}{q-1}.
\end{equation}

The exponent $w$ has the meaning that only moments $\langle
|u|^m\rangle$ of the density with $m<(w -1)$ exist.  This sounds like
a severe restriction on the existence of structure functions, but
since $w$ is rather large ($9 < w < 60$ for the Couette-Taylor data),
this effect does not contradict the experimental measurements of
structure functions. Fig.~4 shows the function $w(r,Re)$ determined
from the experimental data. We observe that $w$ exhibits simple
scaling behaviour for medium spatial scales,

\begin{equation}
w(r) =4 \left( \frac{r}{\eta} \right)^\delta, \label{law}
\end{equation}
where the exponent $\delta$ depends weakly 
on the Reynolds number $Re$. We find $\delta =$
0.440, 0.395, 0.360, and 0.326 for $Re =$ 69000, 133000, 266000, and 540000,
respectively.  For $r/\eta <10$, the experimental data saturate at
$w \approx 9$. For $w>60$, fluctuations become large. 

The constant in front of the experimentally determined
power law is found to be $4.0 \pm 0.1$, independent of
the Reynolds number. A simple argument for the value 4
could be as follows. Suppose we could measure the probability
distribution of ideal turbulence in an unperturbed way down to the
Kolmogorov scale $\eta$ for infinite Reynolds number. Assuming
that the scaling law (\ref{law}) remains valid for $Re
\to\infty$ and $r\to\eta$ we obtain at $r=\eta$ the value $w (\eta) =4$, no
matter what $\delta$ is. But for turbulence to make sense at least
the third moment $\langle |u|^3 \rangle$ should exist, since this is
the most fundamental observable related to energy dissipation.
Existence of the third moment at $r=\eta$ is guaranteed if $w =4+ \epsilon$,
where $\epsilon$ is an arbitrarily small positive number. 
This argument suggests that the constant should be 4.

For $Re \to\infty$ the idealized small-scale
turbulence at $r=\eta$ is characterized by the smallest possible
$w$ where it makes sense to speak about energy dissipation (the
third moment). Since $w=4$, the flatness factor $F=\langle
u^4\rangle/\langle u^2\rangle ^2$ as well as all higher moments at
this scale would then diverge for $Re \to \infty$. This is
compatible with experimental observation \cite{sreeni}.

\section*{Acknowledgement}
$\,$

Part of this research was performed during C.B.'s stay at the
Institute for Theoretical Physics, University of California at Santa
Barbara, supported in part by the National Science Foundation under
Grant No. PHY94-07194. C.B. also gratefully acknowledges support by a
Leverhulme Trust Senior Research Fellowship of the Royal Society.  The
research by H.L.S. was supported by the U.S. Office of Naval Research.

\section*{Figure captions}
$\,$

{\bf Fig.~1}. Experimentally measured probability density functions of
the velocity differences for the Couette-Taylor experiment at
$Re=540000$ are compared with the theoretical curves $\tilde{p}(u)$
(for $\alpha =2-q$): (a) Logarithmic plot, (b) Linear plot. The
logarithmic plot is sensitive to the tails, while the linear plot
is sensitive to the vicinity of the maximum. For the experimental
curves, the distances $r/\eta$ (where $\eta$ is the Kolmogorov length
scale) are, from top to bottom: 11.6, 23.1, 46.2, 92.5, 208, 399, 830,
and 14400. For the theoretical curves, the values of the non-extensivity 
parameter $q$ are, from top to bottom:  1.168,
1.150, 1.124, 1.105, 1.084, 1.065, 1.055, and 1.038.  For better
visibility, each distribution in (a) is shifted by -1 unit along the
$y$ axis, and each distribution in (b) is shifted by -0.1 unit along
the $y$ axis.

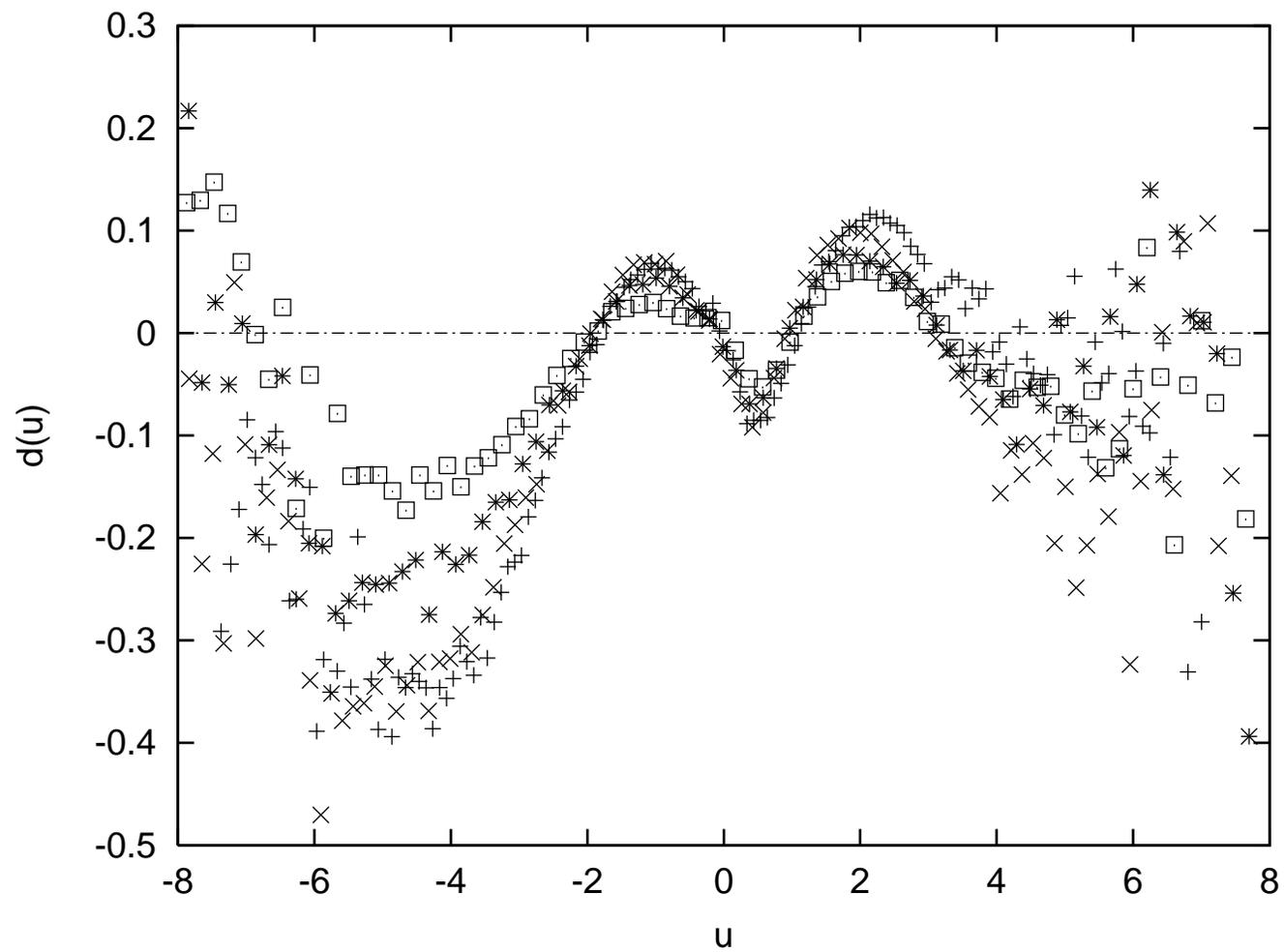
{\bf Fig.~2}. Relative difference $d(u) \equiv
(p_{th}(u)-p_{expt}(u))/p_{th}(u)$ between theoretical and
experimental probability densities at $r/\eta = 12$. The difference
decreases with increasing Reynolds number $Re=$ 69000 $(+)$,
133000 $(\times )$, 266000 $(*)$, and 540000 $(\Box )$.  For each
Reynolds number the best possible fit is used ($q=$ 1.148, 1.159,
1.167, and 1.168, respectively).

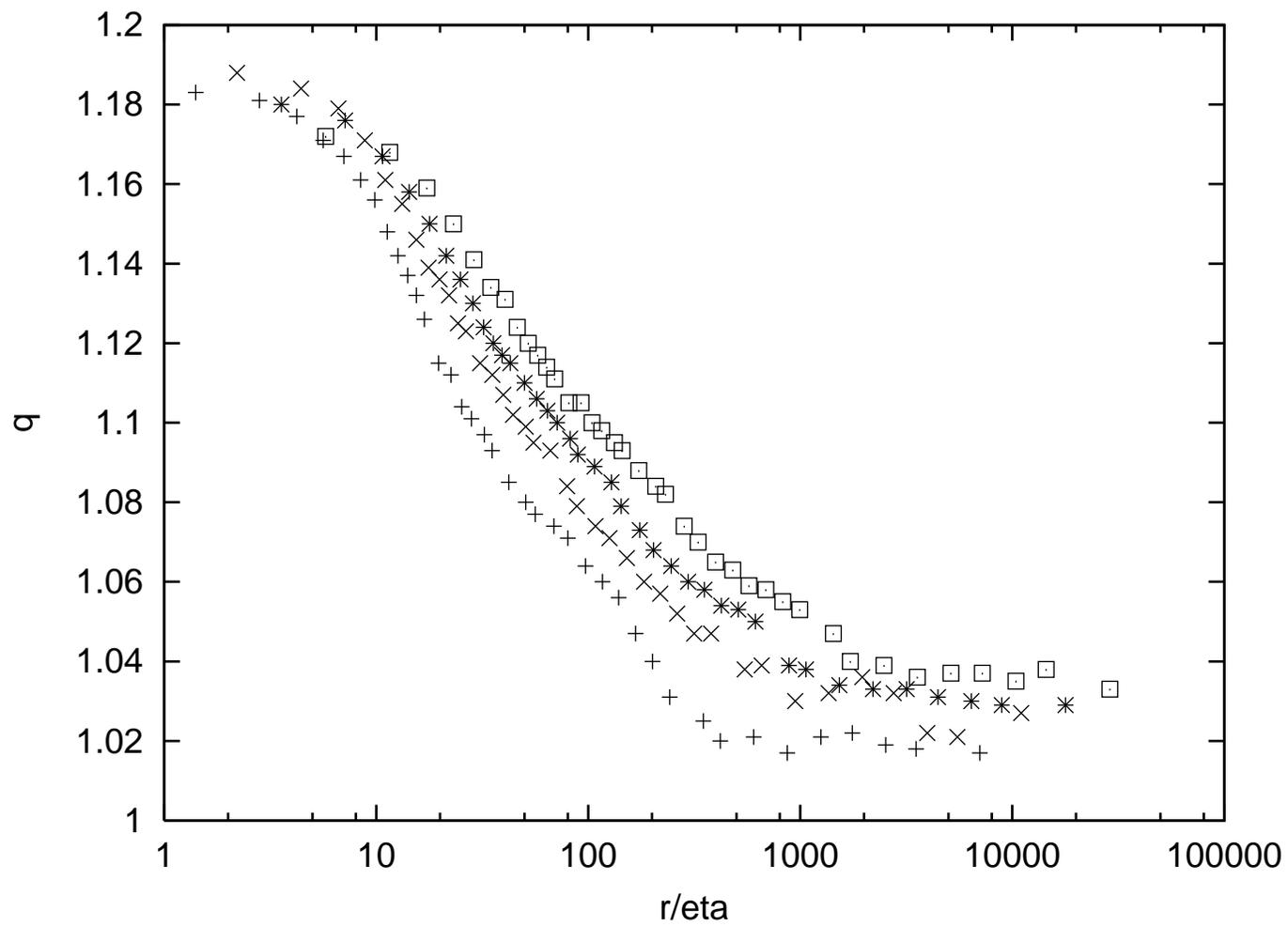
{\bf Fig.~3}. The fitness parameter $q(r,Re )$ deduced from a least
square fit of the velocity data to the theoretical probability
density.  The Reynolds numbers are 69000 $(+)$, 133000 $(\times )$,
266000 $(*)$, and 540000 $(\Box )$.

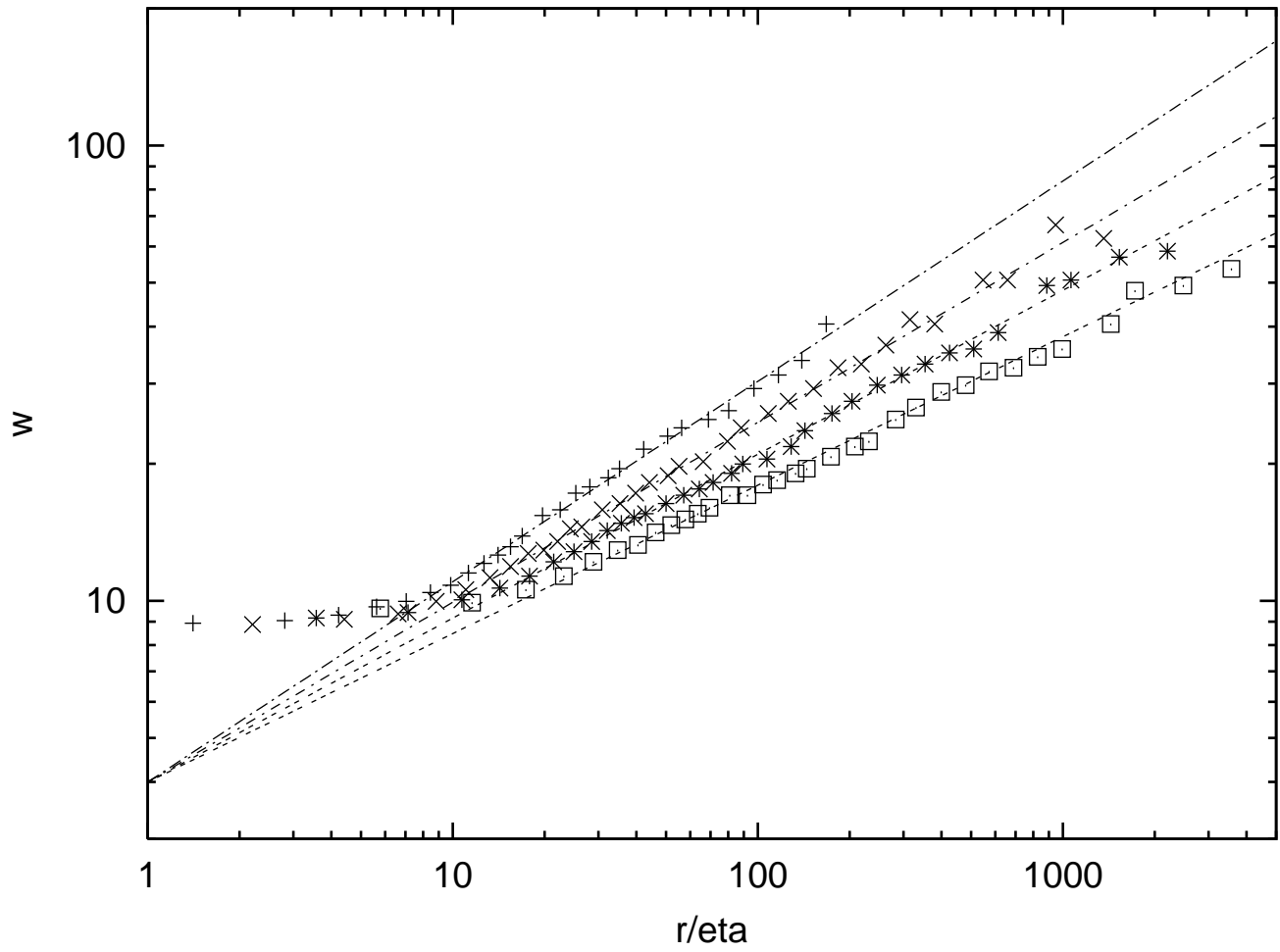
{\bf Fig.~4}. The exponent $w=(4-2q)/(q-1)$ describing the decay rate
of the probability density for large $|u|$.  Scaling behavior of
$w(r)$ is observed for a large range of distances $r$. The Reynolds
numbers are 69000 $(+)$, 133000 $(\times )$, 266000 $(*)$, and 540000
$(\Box )$.  The straight lines correspond to the power law
$w(r)=4(r/\eta)^\delta$ with $\delta =$ 0.440, 0.395, 0.360, and 0.326,
respectively.



\begin{thebibliography}{99}
\bibitem{tsa1} C. Tsallis, J. Stat. Phys. {\bf 52}, 479 (1988)
\bibitem{tsa2} C. Tsallis,  R.S. Mendes and A.R. Plastino, Physica {\bf 261A},
534 (1998)
\bibitem{tsa3} C. Tsallis, Braz. J. Phys. {\bf 29}, 1 (1999)
\bibitem{levy} C. Tsallis, S.V.F. Levy, A.M.C. Souza, R. Maynard,
Phys. Rev. Lett. {\bf 75}, 3589 (1995) 
\bibitem{curado} I. Bediaga, E.M.F. Curado, J. Miranda,
Physica {\bf 286A}, 156 (2000)
\bibitem{parti} C. Beck, Physica {\bf 286A}, 164 (2000)
\bibitem{qua} W.M. Alberico, A. Lavagno, P. Quarati, Eur. Phys. J. {\bf C12},
499 (2000)
\bibitem{turb1} X.-P. Huang and C.F. Driscoll, Phys. Rev. Lett.
{\bf 72}, 2187 (1994)
\bibitem{turb2} B.M. Boghosian, Phys. Rev. {\bf 53E}, 4754 (1996)
\bibitem{turb3} T. Arimitsu and N. Arimitsu, Phys. Rev. {\bf 61E}, 3237
(2000)
\bibitem{hydro} C. Beck, Physica {\bf 277A}, 115 (2000)
\bibitem{web} http://tsallis.cat.cbpf.br/biblio.htm
\bibitem{LS} G.S. Lewis and H.L. Swinney, Phys. Rev. {\bf 59E},
5457 (1999)
\bibitem{hil} A. Hilgers and C. Beck,
Phys. Rev. {\bf 60E}, 5385 (1999)
\bibitem{nel} U. Frisch, P.-L. Sulem, M. Nelkin, J. Fluid Mech.
{\bf 87}, 719 (1978) 
\bibitem{cast} B. Castaing, Y. Gagne, E.J. Hopfinger, Physica
{\bf 46D}, 177 (1990)
\bibitem{hoso} I. Hosokawa, Phys. Rev. {\bf 49E}, 4775 (1994)
\bibitem{mene} C. Meneveau and K.R. Sreenivasan, J. Fluid Mech.
{\bf 224}, 429 (1991)
\bibitem{poly} A. Polyakov, Phys. Rev. {\bf 52E}, 6183 (1995)
\bibitem{detlef} S. Grossmann and D. Lohse, Eur. Phys. Lett. {\bf
21}, 201 (1993)
\bibitem{wang} L.P. Wang, S. Chen, J.G. Brasseur, J.C. Wyngaard,
J. Fluid Mech. {\bf 309}, 113 (1996)
\bibitem{casca} C. Beck, Phys. Rev. {\bf 49E}, 3641 (1994)
\bibitem{sreeni} K.R. Sreenivasan and R.A. Antonia,
Annu. Rev. Fluid Mech. {\bf 29}, 435 (1997)
\end{thebibliography}
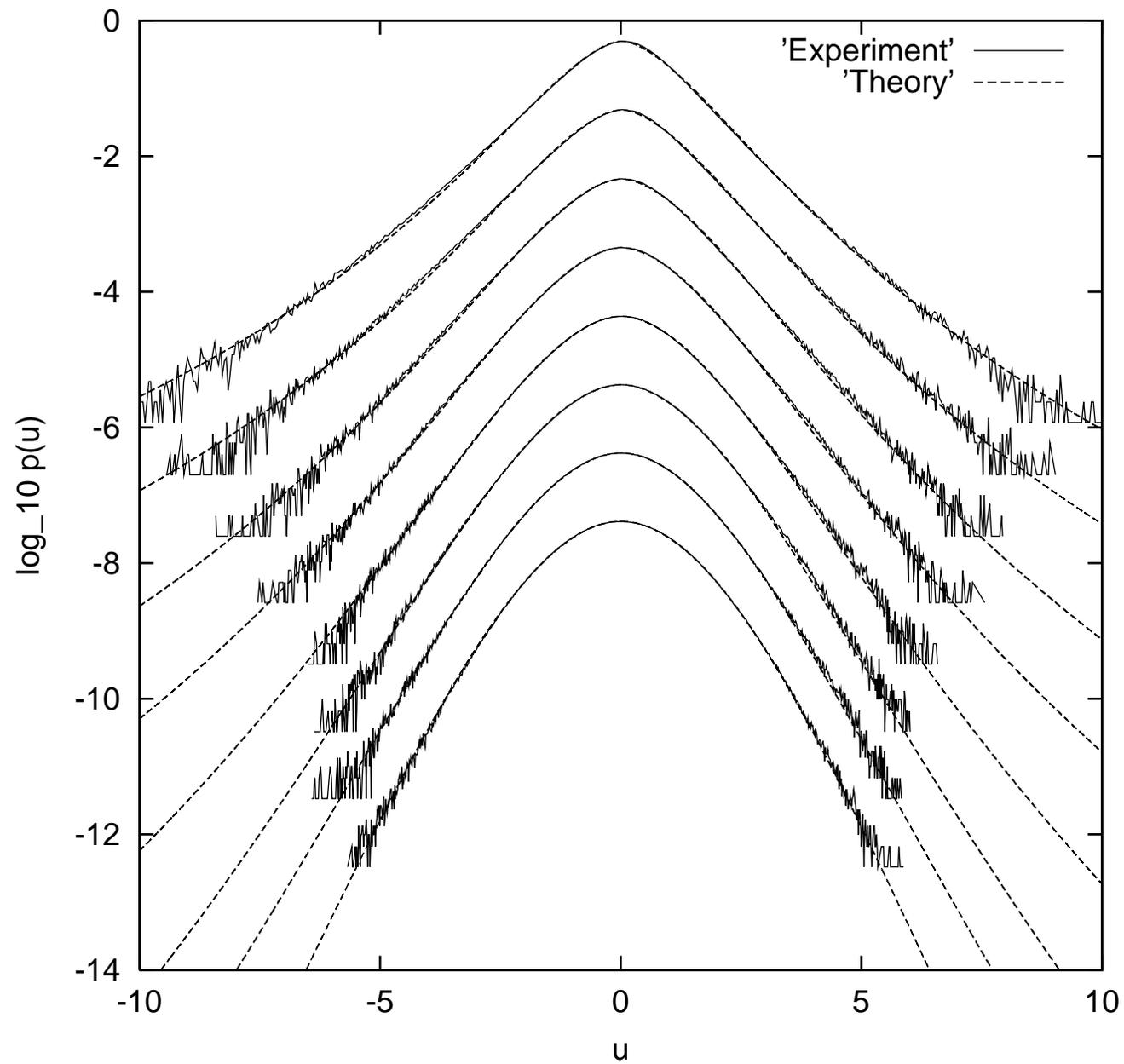
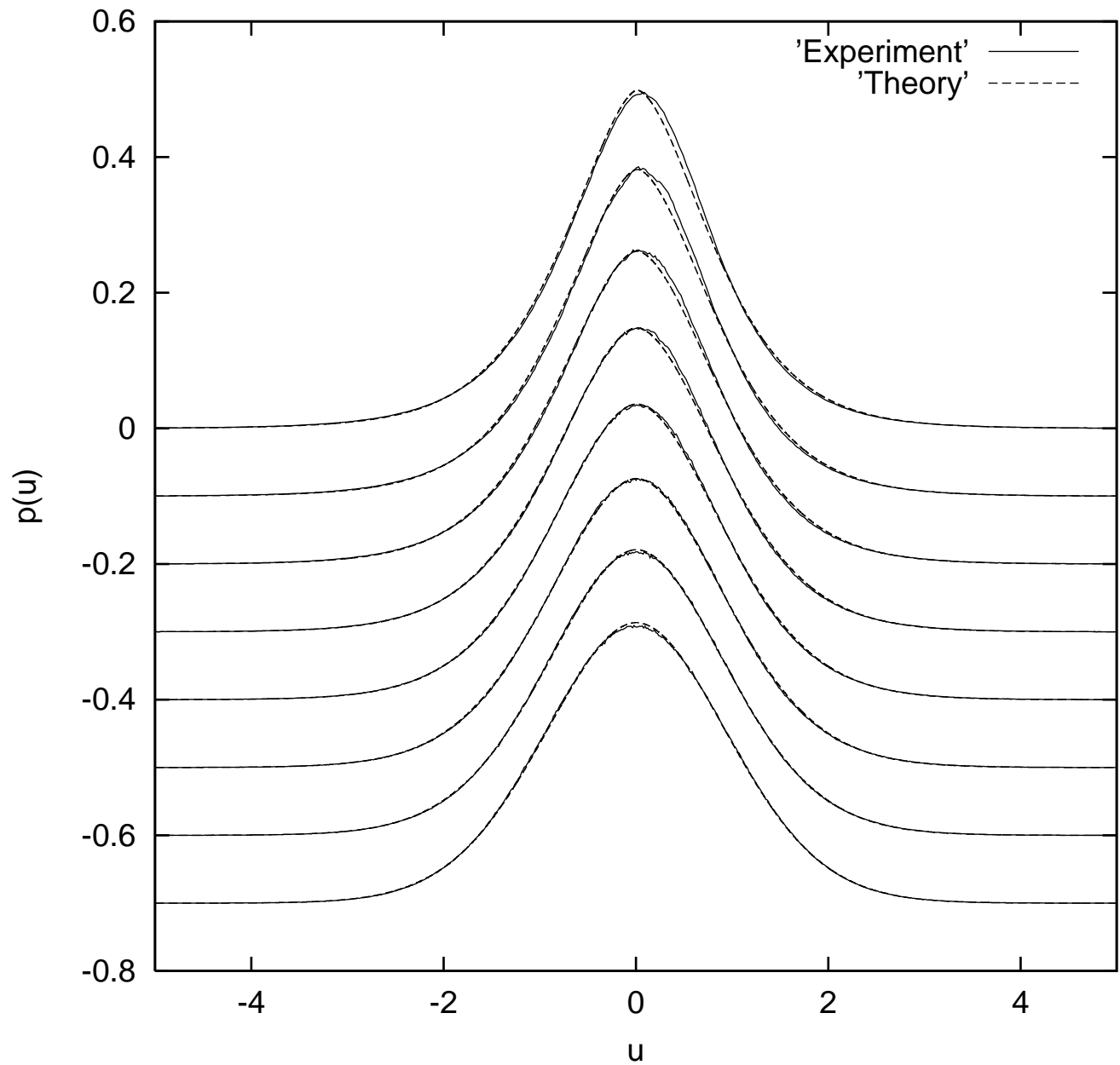
\end{document}